%% file: main.tex
\documentclass{article}
\usepackage{arxiv}

\usepackage[utf8]{inputenc} 
\usepackage[T1]{fontenc}    
\usepackage{url}   
\usepackage{booktabs} 
\usepackage{amsfonts} 
\usepackage{nicefrac} 
\usepackage{microtype}
\usepackage{lipsum}
\usepackage{graphicx}
\graphicspath{ {./figures/} }
\usepackage[version=4]{mhchem} 
\usepackage{tabularx}

\usepackage{siunitx}
\usepackage{multirow}
\usepackage[table]{xcolor}
\usepackage{listings} 
\usepackage{xcolor} 
\usepackage{amsmath}
\usepackage{amssymb}
\usepackage{amsthm}
\usepackage{mathtools}
\usepackage{tikz}
\usetikzlibrary{matrix,chains,positioning,decorations.pathreplacing,arrows}
\usepackage{algorithm}
\usepackage[normalem]{ulem}
\usepackage[noend]{algpseudocode}
\usepackage{filecontents}
\usepackage{xr-hyper}
\usepackage{hyperref} 
\usepackage{cleveref}
\usepackage{comment}
\usepackage{caption}
\usepackage{subcaption}
\usepackage{svg}

\makeatletter
\newcommand*{\addFileDependency}[1]{
  \typeout{(#1)}
  \@addtofilelist{#1}
  \IfFileExists{#1}{}{\ypeout{No file #1.}}
}
\makeatother

\makeatletter
\newcommand*{\addAuxFileDependency}[1]{
  \makeatletter\@input{x#1.tex}\makeatother
}
\makeatother

\small

\numberwithin{equation}{section}
\numberwithin{equation}{subsection}

\algrenewcommand{\algorithmicrequire}{\textbf{Input:}}
\algrenewcommand{\algorithmicensure}{\textbf{Output:}}

\crefname{equation}{Eq.}{Eqs.}
\Crefname{equation}{Equation}{Equations}
\crefname{table}{Table}{Tables}
\Crefname{table}{Table}{Tables}
\crefname{figure}{Fig.}{Figs.}
\Crefname{figure}{Figure}{Figures}

\title{Overcoming the chemical complexity bottleneck in on-the-fly machine learned molecular dynamics simulations}

\author{
  Lucas Timmerman\textsuperscript{a}, Andrew J. Medford\textsuperscript{a}\\
  \texttt{\{ltimmerman3, ajm\}@gatech.edu}\\
  \And
  Shashikant Kumar\textsuperscript{b}, Phanish Suryanarayana\textsuperscript{b,c}\\
  \texttt{shashikant@gatech.edu, phanish.suryanarayana@ce.gatech.edu}\\\\
  \textsuperscript{a}School of Chemical \& Biomolecular Engineering\\
  \textsuperscript{b}School of Civil \& Environmental Engineering\\
  \textsuperscript{c}School of Computational Science and Engineering\\
  Georgia Institute of Technology\\
  Atlanta, Georgia 30332\\
}

\begin{document}

\maketitle

\begin{abstract}
We develop a framework for on-the-fly machine learned force field molecular dynamics simulations based on the multipole featurization scheme that overcomes the bottleneck with the number of chemical elements. Considering bulk systems with up to 6 elements, we demonstrate that the number of density functional theory calls remains approximately independent of the number of chemical elements, in contrast to the increase in the smooth overlap of atomic positions scheme.


\end{abstract}

A multitude of machine learning (ML) models and algorithms have been developed in the past decade to replace and improve first principles or semi-empirical predictions of energies and forces at the atomic scale\cite{Behler2007,Bartok2009,Jinnouchi2019On-the-flyPoints,Drautz2019,Batatia2023AChemistry,Passaro2023ReducingGNNs}. The most accurate of these models rely on graph neural networks or transformer architectures trained on large and carefully curated datasets to learn a map directly from structures to energies or forces\cite{Chen2019GraphCrystals,Chanussot2021,Merchant2023ScalingDiscovery,Tran2023TheElectrocatalysts,Batatia2022MACE:Fields,2023arXiv230612059L,Passaro2023ReducingGNNs,Batatia2023AChemistry}. These models are designed to act as general purpose force fields and have been used successfully to predict the properties of a wide range of materials\cite{Gasteiger2022,Batatia2022MACE:Fields,Lan2022,Sunshine2023ChemicalDensities,PeterKovacs2023MACE-OFF23:Molecules}. However, due to the finite nature of the training sets of these models, edge cases often arise during inference, resulting in unpredictable performance or unstable simulations\cite{Stocker2022}. To address this shortcoming, active learning and fine tuning have been used to dynamically update models\cite{Musielewicz2022FINETUNA:Simulations,Vandermause2022,Podryabinkin2017ActivePotentials,Lysogorskiy2023ActiveModels}. These schemes utilize uncertainty quantification (UQ) methods, such as Bayesian estimates, to determine when a model needs to be updated. This procedure requires interfacing with electronic structure theory software packages to generate additional reference data. UQ methods have also been used to train smaller, data-adaptive models ``on-the-fly'' during a simulation\cite{Jinnouchi2019On-the-flyPoints,Lysogorskiy2023ActiveModels}. Implementations of this approach have been built into popular quantum chemistry packages such as VASP\cite{Jinnouchi2019On-the-flyPoints} and CASTEP\cite{10.1063/5.0155621}, improving the reproducibility of the training procedure and portability of the resulting models. These methods rely on a combination of Bayesian error estimates and heuristics to automatically form training sets, eliminating the burden of manually generating the reference data. In principle, this approach can be used for any chemical system. Unfortunately, the feature-based regression and UQ models used in these schemes cause the cost of both training and inference to increase dramatically with the size of the training set and the number of elements present. Furthermore, the robustness of the on-the-fly training procedure is not well established for complex chemistries.

Due to the inherent trade-offs among the various approaches to machine learning force fields (MLFF), researchers must select from a diverse range of options to identify the model best suited to a given problem. The large, general purpose pretrained models offer a straightforward option since they do not require training and are often designed to work with common scientific packages such as the Atomic Simulation Environment\cite{HjorthLarsen2017} or even have web-based interfaces\cite{Tran2023TheElectrocatalysts}. Beyond accessibility, it is also crucial to evaluate the validity of model predictions with a first-principles  result. This validation is straightforward in the case of large-scale computational screening, where single-point density functional theory (DFT) calculations can be used to validate a subset of candidate materials based on the results of the ML accelerated screening. However, first-principles validation quickly becomes intractable as the quantity of interest moves from DFT energies to properties based on ensemble averages, such as diffusion coefficients or free energies from enhanced sampling\cite{Jakse2013,Bailleul2020AbReaction,Piccini2022AbCatalysis}. This is particularly concerning, as atomistic ML models are known to behave unreliably when making out-of-domain predictions or during extended simulations under reaction conditions\cite{Stocker2022,Fu2022}. As a result, utilizing on-the-fly potentials that quantify uncertainty and dynamically validate and update becomes a favorable strategy for obtaining reliable estimates of properties that require long molecular dynamics (MD) or Monte Carlo simulations.

On-the-fly ML algorithms typically require a close coupling between the ML and DFT codes involved. This can be done by directly integrating the ML with the DFT code, which restricts the complexity of ML algorithms available due to the incompatibility between the lower level languages used to write DFT codes and the more interpretable languages commonly used for cutting edge ML packages\cite{Paszke2019PyTorch:Library,Bezanson2014Julia:Computing,Witt2023ACEpotentials.jl:Expansion}. Alternatively, it is possible to couple DFT codes to advanced ML frameworks through socket interfaces or hybrid language extensions, allowing fine-tuning of pre-trained models on-the-fly\cite{Musielewicz2022FINETUNA:Simulations,Tran2018ActiveEvolution,Shuaibi2021EnablingPriors}. However, this strategy has only emerged recently and substantially increases the complexity of software installation and maintenance. The most prominent on-the-fly implementations are based on legacy smooth overlap of atomic positions (SOAP) chemical descriptors\cite{Bartok2012} and Bayesian linear regression\cite{Jinnouchi2017PredictingAlgorithm}. Existing implementations can be found in the VASP\cite{Jinnouchi2019On-the-flyPoints}, CASTEP\cite{10.1063/5.0155621}, and SPARC\cite{Kumar2024On-the-flyCH} codes. However, the generality of these models is not well established, and the SOAP framework is known to scale poorly to systems with many unique chemical elements due to the increasing computational cost of descriptor computation and inference, as well as the inclusion of redundant information in the descriptor vector\cite{Byggmastar2022,Darby2022,Byggmastar2021ModelingSegregation}.

We aim to reduce the gap between state-of-the-art ML models that work for an arbitrary number of elements but are not always portable or transferable, and the existing SOAP-based on-the-fly force fields that are straightforward to use within DFT codes and work well for simple systems but struggle to scale to systems with many chemical elements. To achieve this goal, we introduce a modified workflow based on the normalized Gaussian multipole (GMP) descriptor\cite{Lei2021}, which shows improved efficiency without compromising performance. See the SI for details on the normalization factor computation. The GMP scheme differs from SOAP in the dependence of the size of the feature vector on the number of unique elements and the formulation of the design matrices. The size of the feature vector for the SOAP chemical descriptor scales quadratically with the number of unique chemical elements, requiring additional computational resources and sometimes causing poor conditioning of the resulting design matrices. Several studies have addressed this issue using compression schemes\cite{Darby2022Tensor-reducedRepresentations,Darby2022}. However, the dimension of the resulting feature vector scales linearly with the number of unique elements, and the procedure to calculate the compressed feature vector increases computational cost and complexity. The GMP-based models overcome this scaling issue by implicitly embedding elemental identity through a Gaussian representation of atomic valence densities, leading to a fixed vector size independent of the number of chemical elements in the system \cite{Liu2020,Lei2021,Shuaibi2023AmpTorch:Quantification}. This lack of explicit elemental dependence leads to denser representations of chemical environments, and allows the use of pooled design matrices that combine the chemical information from all element types via the kernel evaluation. To ensure portability and ease of use, we implemented the GMP-based on-the-fly potentials in the SPARC DFT code\cite{ZHANG2024100649}. The SPARC code has minimal dependencies (MPI\cite{Forum1994MPI:Standard} and BLAS\cite{10.1145/567806.567807}/LAPACK\cite{doi:10.1137/1.9780898719604}/MKL) ensuring ease of compilation and uses a real-space formalism that enables mixed boundary conditions and short wall times, allowing rapid generation of on-the-fly MLFF training data for systems of arbitrary chemical complexity.

To test the implementation, we used a series of bulk metals and alloys with up to six elements. The systems include bulk Al, Ag, Au, Ir, Pd , Pt, and Rh; and alloys of Pt, Ag, Au, Ir, Pd, and Rh, each with 32 atoms in the unit cell. Aluminum was included for benchmarking, as it is a common reference system. The remainder of the elements are of interest in heterogeneous catalysis\cite{Greeley2009,Batchelor2019,Pedersen2020High-EntropyReactions,Greeley2004AlloyPrinciples,Stephens2012UnderstandingAlloys}. 
Single-element systems provide both a benchmark for the performance of the ML algorithms and allow us to compare energies of formation of the complex alloys. All MD simulations were carried out in the isokinetic (NVK) ensemble. The ML formalism is extensible to other ensembles, but the choice of training ensemble determines the range of applicability for the resulting models. Each simulation was run for 10,000 steps (20 ps) with a corresponding ab-initio MD simulation for each system as a ground truth, and we evaluate the models by comparing the total variation distance (TVD)\cite{Gibbs2002OnMetrics} of the pair correlation functions (PCFs). Conceptually, the TVD represents the degree of overlap between the PCFs computed using the MLFF and DFT. Additional details on the TVD calculation are provided in the SI. The use of TVD is inspired by previous work revealing that analysis of PCFs provides the best metric for the stability of machine-learned force fields used for molecular dynamics simulations\cite{Fu2022}. We assess the average TVD over the full trajectory, as well as time-resolved TVDs that provide further insight into stability. We also provide a comparison of the free energies of formation for the alloys to demonstrate a potential application of the GMP MLFF. We compare the common finite displacement (FD) method\cite{Parlinski1997First-principlesZrO2,G.Kresse1995AbGraphite} to MD methods\cite{Schmerler2021Elcorto/pwtools:,Lin2003TheFluids,Minakov2017VibrationalMelting} to obtain thermodynamic corrections to electronic energies.

\begin{figure}[h]
    \centering
    \includegraphics[trim=20 10 20 0,clip=True,keepaspectratio=true,scale=1]{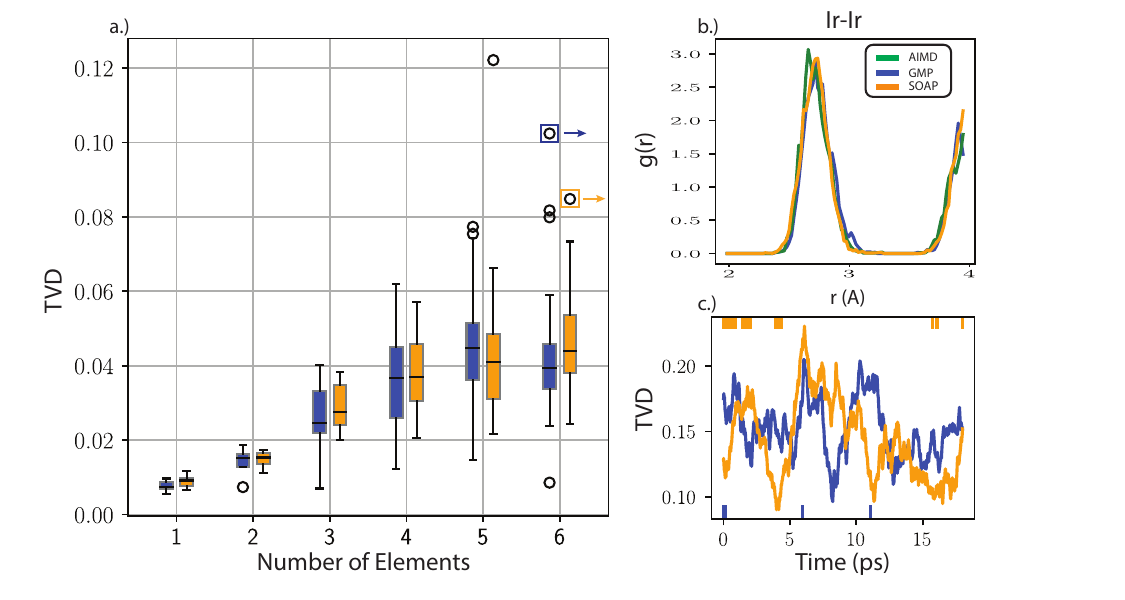}
    \caption{Performance summary for SOAP and GMP models trained on alloys of 1 to 6 elements a.) Mean TVD for all alloys broken down by pariwise interaction for GMP and SOAP in a standard box and whisker plot. The distribution in the case of single-element systems comes from the distribution of all 6 individual single element systems; in all other cases it corresponds to multi-element pair correlation functions.  Open circles correspond to outliers b.) Representative partial PCF. Plots correspond to the outliers highlighted for the 6 element alloy which is the Ir-Ir distribution for both SOAP and GMP c.) Time resolved TVD for SOAP and GMP again for the highlighted outliers. Time resolution was included by computing the TVD for a 2 ps interval incremented by 2 fs across the entire trajectory.}
    \label{fig:alloy4panel}
\end{figure}

Figure \ref{fig:alloy4panel} presents the key accuracy metrics for the AIMD and MLFF models, as well as an illustrative PCF plot to contextualize the TVD metric. We computed the PCFs for all pairwise interactions in each alloy, so the number of distributions scales quadratically with the number of elements. The box and whisker plot in Figure \ref{fig:alloy4panel} a) provides a visualization of the distribution of TVD means broken down by pairwise interactions (for alloys) or by element type (for pure metals). There is a similar increase in the spread of TVD for the GMP and SOAP models associated with an increase in the number of chemical elements, most notably as the number of elements becomes greater than two. We hypothesize that this increasing variance with chemical complexity arises because the total amount of force data per structure remains fixed while the number of unique chemical interactions that must be modeled increases quadratically. It is interesting that the magnitude of this trend is comparable for both the GMP and SOAP models, since the SOAP models explicitly differentiate between unique chemical elements, whereas the GMP models do not. Despite the increase in error for alloys, the median performance for both models appears to stabilize after 4 unique chemical elements. To provide visual context for the TVD metric, we plot the PCF for two outliers of the six-element system in Figure \ref{fig:alloy4panel} b) which corresponds to the same system for GMP and SOAP. Visual inspection of the PCF shows that the ML models accurately reproduce the locations of both major peaks for the system, indicating that despite the relatively large TVD, neither model fails catastrophically. Finally, Figure \ref{fig:alloy4panel} c) shows the time-resolved TVD corresponding to the same Ir-Ir outlier for the 6-element alloy. Although oscillations are clearly present, they do not increase with time and generally remain below a TVD of 0.25 for both models. These findings suggest that the models are stable even for the least accurate systems. We further confirmed the stability of the corresponding GMP models by simulating an additional 100K steps (200 ps) for all systems using the MLFFs in inference-only mode and observed that the formation energies for the alloys shifted by no more than 70 meV relative to the DFT MD values (see convergence table in SI). Overall, these results indicate that the MLFFs are similarly accurate and robust for both SOAP- and GMP-based models.

\begin{figure}[h]
    \centering
    \includegraphics[trim=20 10 0 20,clip=True,keepaspectratio=true,scale=1]{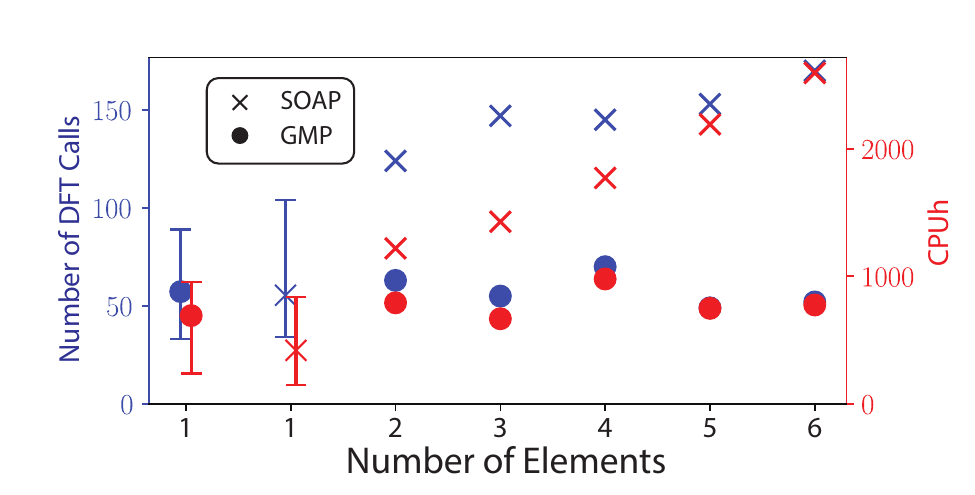}
    \caption{Number of DFT calls (blue, left axis) and cpu times (red, right axis) required to train GMP and SOAP models for all systems. The cpu times include all operations necessary to compute energies and forces using both DFT and the MLFF, as well as time spent training the MLFFs. The error bars for the single element GMP and SOAP systems correspond to the range in number of DFT calls and CPU times among the different pure systems.}
    \label{fig:NDFT}
\end{figure}

Although the accuracy of SOAP and GMP based models are similar, their computational cost differs significantly. In appropriately optimized on-the-fly MLFF codes, the computational cost is dominated by the generation of DFT training data. Figure \ref{fig:NDFT} shows the relationship between the number of chemical elements and the number of DFT calls and the CPU time for both the SOAP and GMP models. Additional efficiency metrics for both models are tabulated in the SI. For SOAP-based models, the number of DFT calls increases sharply when moving from pure metals to alloys and continues to increase as more unique elements are included. The CPU times follow a similar trend and increase steadily as a function of the number of elements. Both observations are attributed to the increase in the size of the SOAP feature vectors with the number of elements. The amount of data needed to train a reliable model generally increases with the dimension of the feature vector due to the curse of dimensionality\cite{Watt2016,Bellman1957DynamicProgramming}, and the CPU time increases because the size of the descriptor vector increases, which adds to the compute and memory requirements for both the calculation of the features and the evaluation of the kernel. In contrast, both the number of DFT calls and the CPU time required for GMP-based models are approximately constant regardless of the number of elements present. This shows that GMP-based on-the-fly models can overcome the elemental scaling bottleneck that will cause SOAP-based models to become unwieldy as the number of elements present increases. We expect to see further improvements as we optimize our implementations of the ML code and extend parallelization of the ML operations. Furthermore, we note that versions of the SOAP descriptor with better elemental scaling \cite{Darby2022,Darby2022Tensor-reducedRepresentations} may lead to improvements similar to the GMP results, though further implementation and testing is needed. 

\begin{figure}[h]
    \centering
    \includegraphics[trim=5 25 0 15,clip=True,keepaspectratio=true,scale=1]{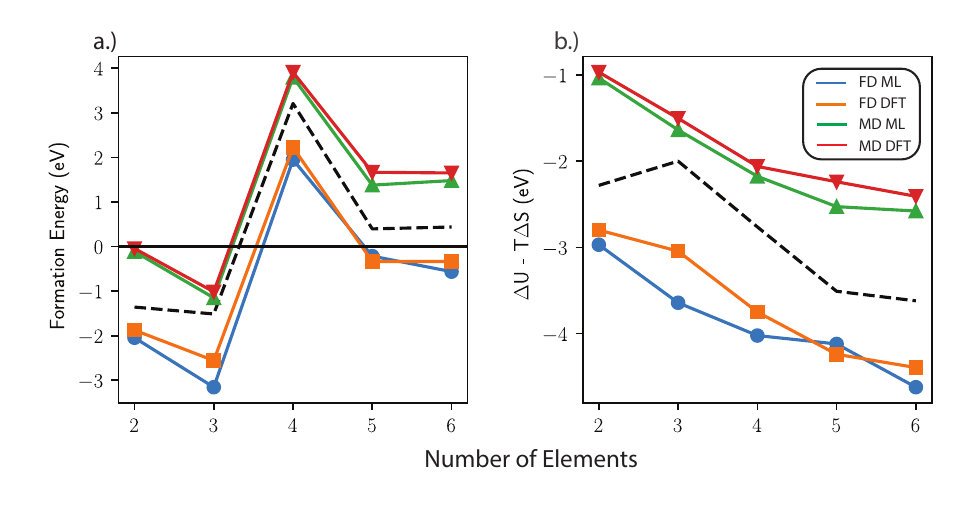}
    \caption{Thermodynamic formation energies for each of the six alloys from the pure bulk components. Each figure contains data for energies computed using phonons extracted directly from MD simulations and the FD method at both the DFT and MLFF levels of theory. The dashed black line in each figure corresponds to the upper bound of the uncertainty associated with low frequency modes and broadening of the PDOS for FD free energy estimates (details in SI). a) Helmholtz free energies of formation with electronic energy included. b) Thermal corrections including vibrational and configurational entropy as well as internal energy.}
    \label{fig:Thermo}
\end{figure}

To demonstrate the utility of the on-the-fly ML models, we use GMP-based models to compute the Helmholtz free energies of formation for each of the five alloys, including internal, vibrational, and configurational contributions. Here, we compare the results obtained using the FD and MD approaches to obtain the phonon density of states. The FD approach is a well-established method that relies on perturbing atoms corresponding to a primitive cell within a supercell structure to obtain an approximate Hessian of a potential energy surface. This Hessian is used to construct the dynamical matrix, from which phonon modes can be extracted\cite{Alfe2009PHON:Method}. The phonon density of states can also be extracted directly from an MD simulation using a Fourier transform of the velocity autocorrelation function\cite{Minakov2017VibrationalMelting,Lin2003TheFluids}. We carry out both procedures at the MLFF and DFT levels of theory. The total free energy for each material was computed from the phonon density of states using ASE's \texttt{CrystalThermo} package. Additional details on these computations and tabulated values can be found in the SI. The corresponding results shown in Fig. \ref{fig:Thermo} show excellent agreement between the DFT and ML formation energies computed using the MD method, while the differences in the ML and DFT FD approximations is substantial ($\sim$0.5 eV) in some cases. However, the FD approximation is plagued by numerical uncertainty associated with the selection of phonon broadening and handling of low-frequency modes. The default options are used for the DFT and ML comparison which includes a broadening of $10^{-3}$ and includes the contributions from imaginary modes by taking the negative square root of the eigenvalue of the Hessian. Alternatively, the black dashed lines in Fig. \ref{fig:Thermo} show the results using the DFT data with slightly different options (broadening of $10^{-4}$ and a low-frequency cutoff equivalent to the largest imaginary mode). The results illustrate that the difference between the DFT and ML models is lower than the numerical uncertainty of the DFT model and highlight the advantage of the MD approach, where these numerical ambiguities are avoided.

Furthermore, it has been shown that the MD method implicitly captures some degree of anharmonicity, which has a large impact on the vibrational contribution to entropy\cite{Ma2015AbOne,Grabowski2019AbAlloys}. This effect is not present in pure metal systems, but it plays a significant role in the vibrational energy of the alloys. The discrepancy between the formation energies computed using the FD and MD methods is dominated by the vibrational term, accounting for $\sim$0.5-1.5 eV of the contribution to the formation energy. Figure \ref{fig:Thermo} b) shows that the difference between the thermal corrections of the FD and MD methods can be as large as $\sim$2 eV. Beyond accounting for anharmonicity, the MD method also provides a straightforward and potentially more efficient procedure for treating thermal corrections in HEAs. 
The number of DFT calls necessary to use the FD method for the 32-atom cell (192) is already greater than the number of DFT calls needed for the MLFF-accelerated MD simulation. Due to the random nature of HEAs, even larger cell sizes may be required for rigorous convergence. The number of DFT calls scales linearly with the number of atoms for the FD approach, while we have demonstrated that the number of DFT calls necessary to train a robust GMP-based MLFF model is approximately independent of the chemical complexity. Thus, the on-the-fly MLFF MD approach will become increasingly more efficient as the cell size increases.

\begin{figure}[h]
    \centering
    \includegraphics[trim=0 35 0 45,clip=True,keepaspectratio=true,scale=0.85]{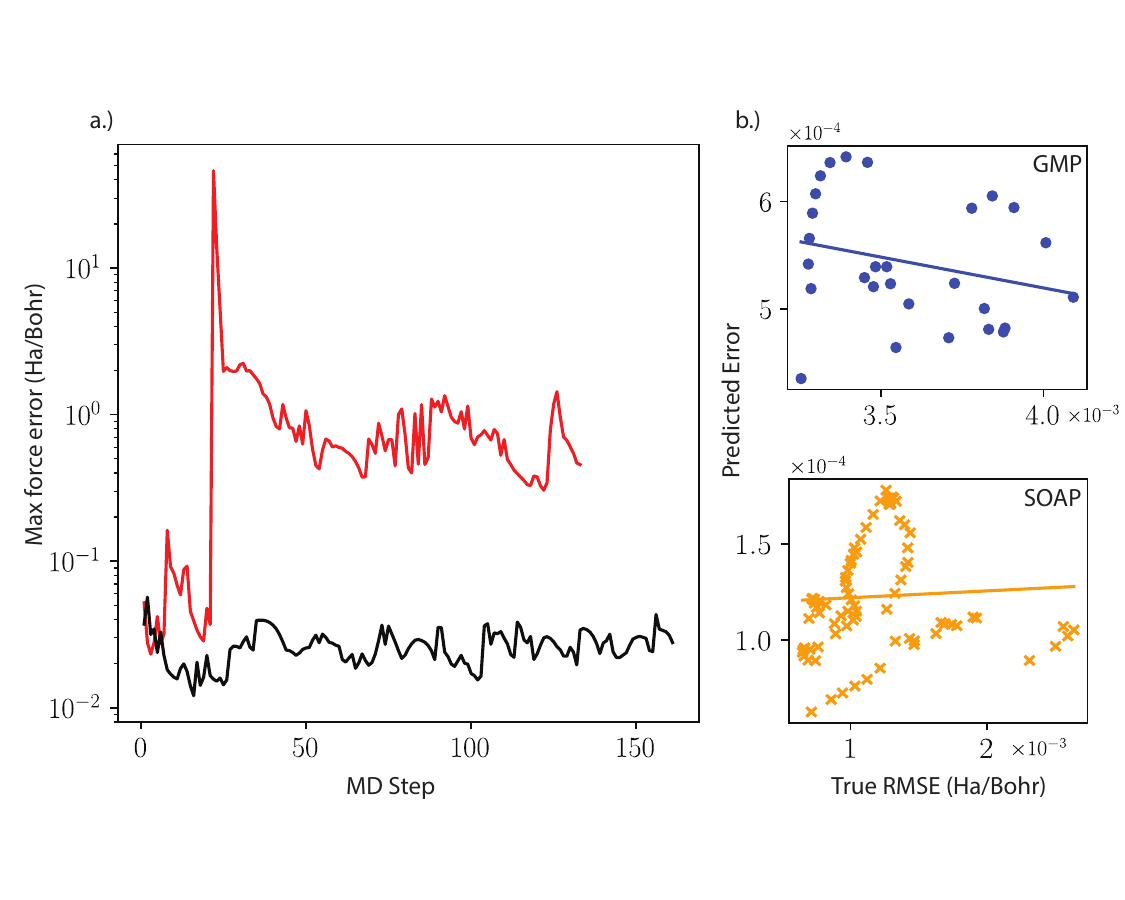}
    \caption{a.) Maximum error of an MD trajectory for the 6 element system system encountered during an on-the-fly run (red) as compared to training from a fixed AIMD trajectory (black). Identical hyperparameters were used in both cases.  b.) Scatter plots showing the lack of correlation between the maximum predicted error and the actual RMSE in the on-the-fly simulation for the 5 element alloy for GMP (top) and for Au for SOAP (bottom). The lines represent linear fits between the true and predicted error.}
    \label{fig:Hyper_UQ}
\end{figure}

The results presented here suggest that on-the-fly potentials with GMP features are a promising strategy for complex chemical systems with many elements. However, some challenges remain. Both SOAP- and GMP-based on-the-fly models require many hyperparameters, some related to how the features are constructed and others related to the uncertainty quantification and the active learning loop. Although it is possible to set reasonable defaults for some of these hyperparameters, others require systematic optimization \cite{Lones2021HowResearchers,Yang2020OnPractice,Bischl2021HyperparameterChallenges} or heuristic tuning. We found that the hyperparameters affecting the ML algorithm shared by GMP and SOAP such as initial training set size and regularization strength dominated the outcomes of on-the-fly runs. We selected the optimal parameters via grid search on the six-element alloy and pure Pt systems. Tuning the descriptor hyperparameters further improved the MLFF performance, although less dramatically. We found that the GMP models improved when we increased the maximum order of the spherical harmonic angular probe and increased the density of radial probes near atom centers. There is a well-established literature regarding the SOAP descriptor hyperparameters, and here we selected parameters conservatively to ensure adequate accuracy. Interestingly, we discovered that using data taken from an AIMD simulation to tune the hyperparameters was not an effective strategy. We attempted to evaluate the model hyperparameters by decoupling the DFT data generation from the training procedure. That is, we utilized the same algorithm to determine when the model needed to be updated, but we drew the first-principles configuration, energy, and force data from a fixed AIMD trajectory. This allowed us to evaluate a far greater set of hyperparameters without requiring additional DFT calls for each run. However, even when the hyperparameters were optimized in this way, the models trained on-the-fly often resulted in unstable simulations as indicated in Figure \ref{fig:Hyper_UQ} a) where the black curve depicts force errors for the above ``offline'' training procedure and the red curve corresponds to the true on-the-fly run. One strategy that was effective was to focus on the most complex systems, since hyperparameters that worked for the 6-element system tended to work well for alloys with fewer elements as well. Nevertheless, hyperparameter optimization required substantial computational and human effort, so establishing more systematic approaches to identifying hyperparameters of on-the-fly ML models for complex chemical systems is an important step to make these approaches more accessible and efficient.

A related challenge is the robustness of on-the-fly models without properly selected hyperparameters. Ideally, the strength of an on-the-fly model is its ability to access the underlying DFT method to ensure reliability. Thus, a user might expect that improper selection of hyperparameters causes a model to be inefficient (i.e. call DFT more often than needed), but should not cause it to yield unphysical results. Unfortunately, this is not the case for the current class of on-the-fly models, where improper hyperparameter selection leads to exploration of highly unphysical portions of phase space resulting in uncontrolled errors, as illustrated in Figure \ref{fig:Hyper_UQ} a). This failure mode is related to the uncertainty quantification and structure of the active learning loop. Typically, the error estimate from the Bayesian regression are assumed to be correlated with the actual root mean squared error\cite{Jinnouchi2019On-the-flyPoints,Zhang2023On-the-flyInterface,Liu2023MolecularDimer,Jaykhedkar2023HowAIMD,Lama2023EnhancedResults}, although the error estimates are known to be poorly calibrated. However, we observe that in some cases the error estimates are not even correlated with the true error. Figure \ref{fig:Hyper_UQ} b) shows the lack of correlation between the Bayesian error estimate and the actual errors for the points where DFT was performed during stable on-the-fly simulations for two different systems with poor correlation (the 5 element system for GMP and Au for SOAP). Pearson's correlation coefficients between the predicted and actual error for all systems can be found in the SI, and in general the correlation is lower for GMP than for SOAP. Despite the poor correlation, these simulations did not result in catastrophic failure, highlighting the lack of direct connection between the quality of UQ estimates and the stability of the simulation. 

The issue of unreliable uncertainty estimates is compounded by the dynamically updated threshold that is often used to overcome the lack of calibration in error estimates \cite{10.1063/5.0180541,Jinnouchi2019On-the-flyPoints}. Once a large error is observed for a training point, the threshold increases, and the model is unable to recover because the DFT calculations are no longer triggered. This can be improved by using hueristics, such as periodically forcing DFT calculations\cite{10.1063/5.0155621}, but this requires additional hyperparameters and can still lead to catastrophic failure between checks. Avoiding catastrophic failures will require the integration of well-calibrated UQ estimates with more robust active learning loops, although the small number of data points and highly correlated nature of MD data make statistically rigorous UQ challenging in the case of on-the-fly MLFF.

Despite these challenges, the results presented here indicate that both SOAP and GMP featurization schemes can be used to construct accurate on-the-fly potentials for systems with up to six unique elements, and the GMP featurization scheme enables development of on-the-fly potentials that are more efficient for many-element systems. We demonstrate that the MLFFs generated on-the-fly are robust enough to be used for simulations at length scales not accessible via AIMD, and demonstrate their utility in computing free energies for complex HEAs. The GMP potentials exhibit favorable scaling in terms of the number of DFT calls necessary to train robust potentials, whereas SOAP models require more DFT data as the number of unique chemical elements increases. The models are implemented in the open-source, portable, and highly parallelized real-space DFT code, SPARC\cite{Xu2021,Ghosh2017SPARC:Clusters,Ghosh2017SPARC:Systems}, so that they can be used by the community and applied to large systems with or without periodic boundary conditions. Future work will focus on improving the efficiency of ML operations, the automation of hyperparameter selection, and the development of more reliable uncertainty quantification and active learning approaches. We expect that this platform will enable wider adoption and application of on-the-fly MLFFs for a wide range of systems relevant to chemistry, materials science, and chemical engineering.

\section{Computational Methods}
\label{s:dft}
We previously adapted the on-the-fly ML algorithm developed by Jinnouchi et. al.\cite{Jinnouchi2019On-the-flyPoints} and inspired by the Gaussian approximation potentials\cite{Bartok2009} to correct orbital free DFT calculations to Kohn-Sham accuracy in the SPARC electronic structure code \cite{ZHANG2024100649, 10.1063/5.0180541}. This implementation has been extended to be compatible with the full Kohn-Sham formalism with the option to use the SOAP\cite{Kumar2024On-the-flyCH} or GMP descriptors with all on-the-fly functionality implemented in a development branch. The initial training set size and regularization strength for the ML models were systematically optimized using a grid search routine on the 6 element alloy and pure Pt. Descriptor parameters were either taken from the literature or tweaked heuristically to improve accuracy. We used 2x2x2 bulk super cells with 32 atoms for all systems. We utilized the PBE exchange correlation functional with the D3 correction scheme of Grimme\cite{Grimme2010AH-Pu}. The k-point density and mesh spacing were adjusted until the DFT energy was converged to at least 10$^{-3}$ Ha/atom. 
The SPMS ONCV pseudopotentials with non-linear core corrections were used to treat core electrons\cite{Shojaei2023SoftOptimization,Hamann2013OptimizedPseudopotentials}. The calculations were performed with periodic boundary conditions in all principal directions. All systems were treated as spin-paired since none of the metals considered are magnetic \cite{Lopanitsyna2022ModelingCompression}.
We validated the accuracy of our AIMD simulations using the blocking method to quantify the blocked standard error for the distance associated with the dominant peak in each all-atom PCF\cite{Flyvbjerg1989ErrorData,Grossfield2009ChapterSimulations}. The standard error did not exceed 10$^{-2}$ Bohr for any of the simulations indicating that the simulations were valid. Free energies were computed using the ASE \texttt{CrystalThermo} package for both the FD and MD methods. The phonon denisty of states were extracted using the ASE \texttt{Phonons} module for the FD method with both DFT and ML calculators for force estimations. The phonon density of states were extracted from MD simulations using the \texttt{pwtools} python package. Additional details are provided in the SI.

\section{Acknowledgements}

The authors gratefully acknowledge the support of the U.S. Department of Energy, Office of Science under grant DE-SC0023445. This research was also supported by the supercomputing infrastructure provided by Partnership for an Advanced Computing Environment (PACE) through its Hive (U.S. National Science Foundation through grant MRI-1828187) and Phoenix clusters at Georgia Institute of Technology, Atlanta, Georgia. The views and conclusions contained in this document are those of the authors and should not be interpreted as representing the official policies, either expressed or implied, of the Department of Energy, or the U.S. Government.

\bibliography{main}
\bibliographystyle{unsrt} 

\include{SI}

\end{document}

%% file: SI.tex
\renewcommand{\thefigure}{S\arabic{figure}}
\setcounter{figure}{0} 
\renewcommand{\thetable}{S\arabic{table}}

\section{Supporting Information}
All data and scripts can be found at the Zenodo repo \href{https://zenodo.org/doi/10.5281/zenodo.11217973}{here}.
\subsection{Normalized GMP Descriptor}
The GMP descriptor, originally published by Lei. et. al., is a universal mapping from Cartesian space to chemical descriptor space that embeds elemental identity via Gaussian approximations to pseudopotential valence electron densities. The original formulation of the descriptor was not suitable for on-the-fly kernel based ML as the units of the descriptors and hence magnitudes varied wildly. To overcome this issue, we apply a simple normalization factor computed at the beginning of every simulation based on the selected radial $g_n(r)$ and angular $Y_m$ probe functions:

\begin{align*}
    A_{n,m} & = \frac{1}{\sqrt{\iiint g_n^2(r) Y_m^2(\theta,\phi) dr d\theta d\phi}}
\end{align*}

where $g_n(r)$ represents a Gaussian with a specified standard deviation of $n$ and $Y_m$ represents the collection of Maxwell Cartesian Spherical Harmonics of order $m$. This scaling factor ensures uniformity in units across different angular channels and forces that magnitude of higher order descriptors to 0 as in traditional multipole expansions. This step allowed us to use traditional kernel methods to evaluate the similarity between chemical environments represented by the GMP descriptors.

\subsection{Additional Computational Details}
Primitive \verb|fcc| unit cells were constructed for Ag, Al, Au, Ir, Pd, Pt, and Rh with initial guesses for lattice parameters taken from the Materials Project. The lattice constants were determined utilizing the \texttt{fmin} function from \texttt{sklearn} to minimize the energy as a function of lattice constant. 
Super cells were created from relaxed primitives by repeating cells. The alloys were formed by randomly replacing atoms in a Pt super cell with $\sim$equimolar amounts of the alloyed elements. The mixing was done element-wise proceeding alphabetically, resulting in five random bulk alloys (PtAg, PtAgAu, PtAgAuIr, PtAgAuIrPd, PtAgAuIrPdRh). Initial guesses for the lattice parameters of the super cells were formed via composition-weighted averages of the lattice constants for the constituent elements. Relaxations on the cell and positions were carried out iteratively until both the maximum pressure in the cell fell below 1E-02 GPa and the maximum force component fell below 6E-04 Ha/Bohr. Relaxed cells were used as starting points for all MD simulations, both AIMD and on-the-fly. We used the isokinetic (NVK) ensemble with a Gaussian thermostat to maintain the simulation temperature at 500K. A time step of 2 fs was used for integration. The simulations were run until 10,000 integration steps were completed, resulting in a simulation length of 20 ps. This provided an adequate balance between a desire to minimize computational cost and the need to robustly characterize the stability of the simulations. 
We ran all calculations with a k-point density of 32 \r{A} which corresponds to a k-point grid of (8,8,8) for the primitive fcc cells and (4,4,4) for the supercells. We employed mesh spacing of 0.25 Bohr which is equivalent to a plane-wave cutoff of $\sim$1340 eV. The SCF tolerance was set to achieve at least 1E-4 Ha/Bohr accuracy in atomic forces for all relaxations and MD simulations.

We defined the TVD for integral normalized PCFs as:
\begin{equation*}
    TVD = \frac{1}{2} \int_0^{R_{max}} |g_{DFT}(r) - g_{ML}(r)| dr
\end{equation*}
where $g(r)$ represents the integral normalized PCF for a given interaction and the integration domain is radial distance. Normalizing the PCFs by their integral area allows us to treat them as probability density functions that describe the probability of finding a given pair of atoms at a specified radial distance from each other. This allows for a straightforward application of the TVD metric.

We define the formation energy of the alloys as the energy of the alloy minus the composition weighted sum of the constituent components:
\begin{equation*}
    E_f = E_{alloy} - \sum_i^{N_{types}} n_i E_i
\end{equation*}
where $E_i$ corresponds to the per atom energy of species $i$ in its reference state and $n_i$ corresponds to the number of atoms of species $i$ in the alloy. We include electronic, internal, vibrational, and configurational contributions in our calculation. The configurational entropy was computed assuming an ideal solid solution:
\begin{equation*}
    S_{conf} = -k_B \sum_i^{N_types} x_i ln x_i
\end{equation*}
This term is 0 in the case of pure bulk components thereby acting as a strong driving force for the formation of the HEAs. The computation of electronic, internal, and vibrational contributions were computed differently for the FD and MD approaches. For both methods, the electronic contribution was taken as the fully relaxed or ground state energy computed using DFT. The phonon density of the states was calculated using both DFT and trained ML models to assemble and diagonalize the dynamical matrix from a numerical approximation of the PES Hessian using the built-in ASE \texttt{Phonons} module. For the MD method, the phonon density of states was extracted directly from the simulation via the Fourier transform of the velocity autocorrelation function using the \texttt{pwtools} python package. In both the FD and MD method, the internal and vibrational energies were calculated using the ASE \texttt{CrystalThermo} package. Note that the ASE module automatically normalizes the phonon density of states to match the expected vibrational degrees of freedom in the system. The density of states extracted from the MD simulations was manually normalized. 

\subsection{Convergence of Formation Energy}
Table \ref{tab:long_thermo_Convergence} contains the convergence information for the extended ML-only runs. These values represent the deviation of the formation energy of each alloy from the ML MD value computed after 20 ps.

\begin{table}[h]
    \centering
    \resizebox{\textwidth}{!}{%
    \begin{tabular}{lrrrrrr}
    \toprule
    Formation Energy Convergence & 9.777 & 69.842 & 10.689 & 12.101 & 3.624 & meV\\
    \bottomrule
    \end{tabular}
    }
    \caption{Difference between reference MD ML formation energies for the 20 ps run and MD ML formation energies computed running the simulation in predict only mode for an additional 200 ps.}
    \label{tab:long_thermo_Convergence}
\end{table}

\subsection{Efficiency Data}
Table \ref{tab:eff_tab_alloy} contains a summary of efficiency data for the SOAP and GMP models during on-the-fly training on the alloys. In every case, the efficiency of the on-the-fly procedure favors the GMP models. The GMP models require anywhere from 2-3x fewer DFT calls, 4-5x fewer training atoms, and less overall walltime. The simulation time here is dominated by DFT calls. The amount of walltime spent in ML operations is occasionally larger for GMP, but this inverts as the number of elements in the systems increases due to an increase in the walltime for SOAP models and no change in the GMP walltimes. 

\begin{table}[h!]
    \centering
    \resizebox{\textwidth}{!}{%
    \begin{tabular}{llrrrrr}
    \toprule
                 &      &  No. of KS &  WTime DFT (h) &  WTime MLFF (h) &  Total WTime (h) &  Columns \\
    Material System & Method &  Steps Performed                           &                &                 &                  &          \\
    \midrule
    \multirow{2}{*}{PtAg} & GMP &                         63 &           2.31 &            0.99 &             3.30 &       47 \\
                 & SOAP &                        124 &           4.67 &            0.41 &             5.08 &      203 \\
    \cline{1-7}
    \multirow{2}{*}{AuPtAg} & GMP &                         55 &           1.78 &            1.00 &             2.78 &       39 \\
                 & SOAP &                        147 &           5.13 &            0.82 &             5.95 &      209 \\
    \cline{1-7}
    \multirow{2}{*}{IrPtAgAu} & GMP &                         70 &           3.02 &            1.06 &             4.08 &       55 \\
                 & SOAP &                        145 &           6.31 &            1.07 &             7.38 &      227 \\
    \cline{1-7}
    \multirow{2}{*}{PdPtAgAuIr} & GMP &                         49 &           2.11 &            1.01 &             3.12 &       49 \\
                 & SOAP &                        153 &           6.55 &            2.58 &             9.13 &      230 \\
    \cline{1-7}
    \multirow{2}{*}{RhPtAgAuIrPd} & GMP &                         52 &           2.24 &            0.99 &             3.23 &       52 \\
                 & SOAP &                        170 &           8.45 &            2.36 &            10.81 &      276 \\
    \bottomrule
    \end{tabular}
    }
    \caption{Selected metrics for assessing model efficiency for GMP and SOAP simulations on alloys. The timings are approximate and correspond to wall times. CPU hours may be computed by multiplying the wall times by 240. The "Columns" metric refers to the number of columns in the design matrix. The number of rows corresponds direcly to the number of KS steps.}
    \label{tab:eff_tab_alloy}
\end{table}

Table \ref{tab:eff_tab_sm} contains the efficiency data for the single element systems. These results differ from the alloy systems in a couple of ways. First, the number of DFT calls is similar during the on-the-fly simulation for GMP and SOAP models. Second, the wall time spent on MLFF operations heavily favors the SOAP based models in all but one case. Interestingly, the number of columns in the GMP single element models is on par with the the number of columns in the alloy models which suggests that the GMP descriptor may be more information dense than SOAP. The SOAP models, on the other hand, contain nearly 2x fewer columns for the single element systems than the alloys. The total wall times for the GMP models are comparable to the alloy systems, whereas the wall times for the SOAP runs are greatly reduced. 

\begin{table}[h!]
    \centering
    \resizebox{\textwidth}{!}{%
    \begin{tabular}{llrrrrr}
    \toprule
       &      &  No. of KS &  WTime DFT (h) &  WTime MLFF (h) &  Total WTime (h) &  Columns \\
    Material System & Method & Steps Performed                           &                &                 &                  &          \\
    \midrule
    \multirow{2}{*}{Al} & GMP &                         33 &           0.25 &            0.74 &             0.99 &       33 \\
       & SOAP &                         52 &           0.48 &            0.13 &             0.61 &      129 \\
    \cline{1-7}
    \multirow{2}{*}{Ag} & GMP &                         60 &           1.80 &            0.72 &             2.52 &       31 \\
       & SOAP &                         59 &           1.64 &            0.09 &             1.73 &       81 \\
    \cline{1-7}
    \multirow{2}{*}{Au} & GMP &                         89 &           2.90 &            1.07 &             3.97 &       29 \\
       & SOAP &                        104 &           3.40 &            0.08 &             3.48 &       81 \\
    \cline{1-7}
    \multirow{2}{*}{Ir} & GMP &                         50 &           1.81 &            1.35 &             3.16 &       29 \\
       & SOAP &                         34 &           1.07 &            0.07 &             1.14 &       62 \\
    \cline{1-7}
    \multirow{2}{*}{Pd} & GMP &                         62 &           2.46 &            0.86 &             3.32 &       36 \\
       & SOAP &                         53 &           1.98 &            0.09 &             2.07 &       87 \\
    \cline{1-7}
    \multirow{2}{*}{Pt} & GMP &                         64 &           2.35 &            1.24 &             3.59 &       28 \\
       & SOAP &                         46 &           1.59 &            0.07 &             1.66 &       64 \\
    \cline{1-7}
    \multirow{2}{*}{Rh} & GMP &                         43 &           1.75 &            0.88 &             2.63 &       34 \\
       & SOAP &                         40 &           1.45 &            0.09 &             1.54 &       82 \\
    \bottomrule
    \end{tabular}
    }
    \caption{Selected metrics for assessing model efficiency for GMP and SOAP simulations on single element bulk structures. The metrics are the same as in Table \ref{tab:eff_tab_alloy}}
    \label{tab:eff_tab_sm}
\end{table}

\subsection{Tabulated Values of Energies}

Here, we include tabulated values of internal energies and entropies (Table \ref{tab:U_TDelS}) and formation energies (Table \ref{tab:form_energy}) corresponding to the figures of the main text for reference, as well as tabulated values of the contributions due to configurational entropy (\ref{tab:Config_S}. 

\begin{table}[h]
    \centering
    \resizebox{\textwidth}{!}{%
    \begin{tabular}{lrrrrrr}
    \toprule
    Number of Elements & 2 & 3 & 4 & 5 & 6 & eV\\
    \midrule
    FD ML & -2.968360 & -3.641422 & -4.021960 & -4.119515 & -4.621012 & \\
    FD DFT & -2.797997 & -3.043473 & -3.748951 & -4.239140 & -4.393300 & \\
    FD Uncertainty Bound & -2.279862 & -1.998956 & -2.761698 & -3.508418 & -3.619274 & \\
    MD ML & -1.034114 & -1.636513 & -2.174891 & -2.526585 & -2.577619 & \\
    MD DFT & -0.966875 & -1.501839 & -2.060034 & -2.238274 & -2.406698 & \\
    \bottomrule
    \end{tabular}
    }
    \caption{Differences in internal energy and entropy at 500 K for the formation of the alloy from its constituent elements. Includes the configurational entropy of the alloy as an ideal solid solution}
    \label{tab:U_TDelS}
\end{table}

\begin{table}[h]
    \centering
    \resizebox{\textwidth}{!}{%
    \begin{tabular}{lrrrrrr}
    \toprule
    Number of Elements & 2 & 3 & 4 & 5 & 6 & eV\\
    \midrule
    FD ML & -2.045798 & -3.154633 & 1.950198 & -0.214359 & -0.562661 & \\
    FD DFT & -1.875435 & -2.556684 & 2.223207 & -0.333984 & -0.334949 & \\
    FD Uncertainty Bound & -1.952282 & -2.580561 & 2.174884 & -0.450343 & -0.371092 & \\
    MD ML & -0.111552 & -1.149723 & 3.797267 & 1.378571 & 1.480732 & \\
    MD DFT & -0.044313 & -1.015050 & 3.912124 & 1.666882 & 1.651653 & \\
    \bottomrule
    \end{tabular}
    }
    \caption{Formation Energies. Includes electronic energy as well. Equivalent to a Helmholtz free energy of formation.}
    \label{tab:form_energy}
\end{table}

\begin{table}[h!]
    \centering
    \resizebox{\textwidth}{!}{%
    \begin{tabular}{lrrrrrr}
    \toprule
    Number of Elements & 2 & 3 & 4 & 5 & 6 & eV\\
    \midrule
    Config S & 0.95569 & 1.51337 & 1.91138 & 2.21504 & 2.46514 & eV\\
    \bottomrule
    \end{tabular}
    }
    \caption{Configurational entropy corresponding to an ideal solid solution for each of the alloys considered here. The numbers were computed by assuming that 1 mol of alloy corresponds to $N_{Av}$ atoms in the ratio equivalent to that represented in our 32 atom supercells.}
    \label{tab:Config_S}
\end{table}

\subsection{Pearson Correlation Coefficients for UQ}
\begin{figure}[h]
    \centering
    \includegraphics[trim=0 0 0 0,clip=True,keepaspectratio=true,scale=0.5]{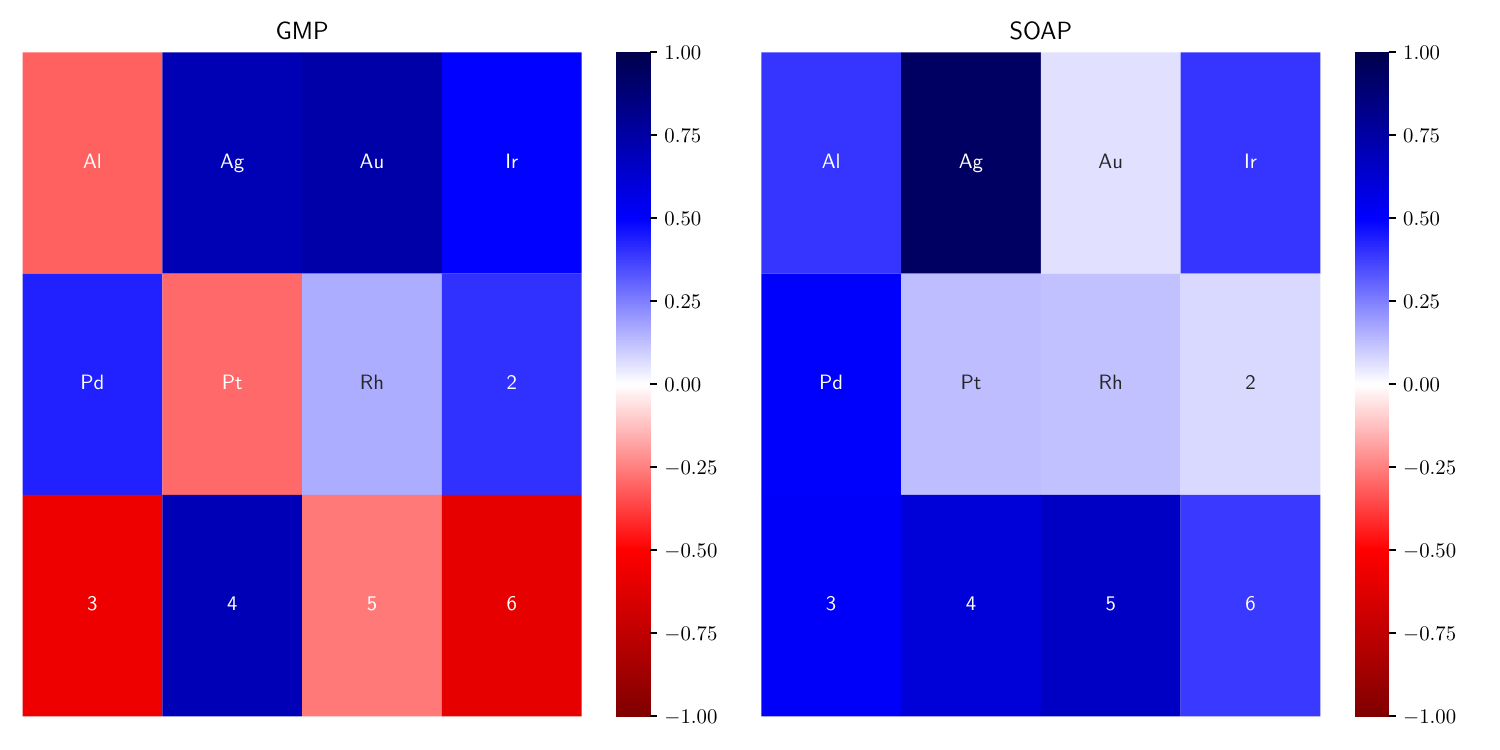}
    \caption{Heatmap of correlation coefficients for the maximum predicted error and the actual RMS error in the system.}
    \label{fig:corr_comb}
\end{figure}

In this work, we utilized the maximum diagonal component of the Bayesian covariance matrix as the uncertainty quantification metric. Figure \ref{fig:corr_comb} presents the Pearson's correlation coefficient between the maximum Bayesian error estimate for the forces and the actual RMS force errors for all on-the-fly simulations. It is clear that there is great disparity in the quality of the Bayesian estimate as a probe for the actual error in the system. Several features are of special interest. First, there do not appear to be any clear trends for the quality of the estimate. There is variation both within the single-element systems and as new species are introduced via the alloys. Second, the absolute values of the correlations coefficients range from nearly 0 to 1 for SOAP, whereas the range for GMP is -0.5 to 0.7. It is rare that the correlation exceeds 0.6 regardless of the model type used.

